\documentclass[letterpaper,12pt,oneside,onecolumn]{article} 
\usepackage{amssymb,amsmath}
\usepackage{txfonts}
\usepackage{graphicx}

\begin{document}

\title{Dynamics and thermalization in correlated 
         one-dimensional lattice systems}

\author{Marcos Rigol\footnote{Present address: Department of Physics, 
The Pennsylvania State University, University Park, Pennsylvania 16802, USA}
\\ Department of Physics, Georgetown University, \\ Washington, DC~20057, USA}

\date{}

\maketitle

\begin{abstract}
We review exact approaches and recent results related to the relaxation dynamics and description 
after relaxation of various one-dimensional lattice systems of hard-core bosons after a sudden 
quench. We first analyze the integrable case, where the combination of analytical 
insights and computational techniques enable one to study large system sizes. 
Thermalization does not occur in this regime. However, after relaxation, observables can 
be described by a generalization of the Gibbs ensemble. We then utilize full exact 
diagonalization to study what happens as integrability is broken. We show that thermalization 
does occur in finite nonintegrable systems provided they are sufficiently far away from the 
integrable point. We argue that the onset of thermalization can be understood in terms
of the eigenstate thermalization hypothesis.
\end{abstract}

\section{Introduction}
Understanding how statistical properties emerge from microscopic models of many-particle 
systems is of fundamental interest in several fields in physics. 
This topic has been extensively studied in the context of classical systems. We know that 
if we perturb a generic isolated gas in many different ways, it will still relax to a unique 
(Maxwell) velocity distribution determined by its energy. This universal 
behavior (thermalization) has been understood in terms of dynamical chaos, namely, the nonlinear 
equations that drive the dynamics ensure that the system explores ergodically all the available 
phase space \cite{gallavotti_book_99}. However, there is a class of models, known as integrable 
models, for which the presence of a full set of conserved quantities precludes thermalization.
In this case, dictated by the initial conditions, the dynamics is restricted to a limited region 
of phase space. More than fifty years ago, Fermi, Pasta, and Ulam (FPU) \cite{fermi_pasta_55} 
set up one of the first numerical experiments to study how thermalization takes place in a 
one-dimensional (1d) lattice of harmonic oscillators once nonlinear couplings were added. 
No signs of ergodicity were found. Those unexpected results led to intensive research 
\cite{campbell_rosenau_05} and ultimately to the development of modern chaos theory 
\cite{rasetti_book_86}.

Recent advances in cooling and trapping atomic gases has led to increased interest in 
understanding what happens in the quantum case. In those experiments, the high degree of 
isolation, combined with the possibility of controlling interactions and the effective 
dimensionality of the gas, has allowed experimentalists to realize 
\cite{kinoshita_wenger_04,paredes_widera_04,kinoshita_wenger_05} and explore the dynamics
\cite{kinoshita_wenger_06,hofferberth_lesanovsky_07} of nearly integrable 1d systems.
Thermalization was not observed in one of the experiments \cite{kinoshita_wenger_06} but was 
indirectly confirmed in the other \cite{hofferberth_lesanovsky_07}. These results have 
motivated intense theoretical research on the dynamics and thermalization 
of isolated quantum systems after a sudden quench, both in the integrable 
\cite{rigol_dunjko_07,cazalilla_06,rigol_muramatsu_06,calabrese_cardy_07,cramer_dawson_08,
barthel_schollwock_08,kollar_eckstein_08,cramer_flesch_08b,rossini_silva_09,iucci_cazalilla_09,
mossel_caux_10,barmettler_punk_10,fioretto_mussardo_10,cassidy_clark_11} and nonintegrable \cite{kollath_lauchli_07,
manmana_wessel_07,rigol_dunjko_08,moeckel_kehrein_08,reimann_08,roux_09,eckstein_kollar_09,
rigol_09a,rigol_09b,moeckel_kehrein_09,santos_rigol_10STATc,roux_10,reimann_10,rigol_santos_10} regimes.

Here, we review results for 1d systems of hard-core bosons (HCBs) on a lattice. 
We show that 
thermalization does not occur (in general) when the system is integrable. However, observables 
after relaxation can be described by a generalization of the Gibbs ensemble 
\cite{rigol_dunjko_07,rigol_muramatsu_06,cassidy_clark_11}. As integrability is broken, thermalization does take 
place \cite{rigol_09a,rigol_09b,rigol_santos_10}, and is shown to follow after the eigenstate 
thermalization hypothesis \cite{rigol_dunjko_08,deutsch_91,srednicki_94}.

\section{Methodology} 
The HCB Hamiltonian of interest reads
\begin{equation}
\begin{split}
\hat{H}_\textrm{HCB}=&\sum_{j=1}^{L} \left[
 - J \left( \hat{b}_{j}^{\dagger} \hat{b}_{j+1} + \textrm{H.c.} \right)
 + V \left( \hat{n}_{j}-\dfrac{1}{2}\right)\left( \hat{n}_{j+1}-\dfrac{1}{2}\right)\right]
 + \sum_{j=1}^{L}V^{\textrm{ext}}_{j} \hat{n}_{j}   \\ 
& +\sum_{j=1}^{L}\left[- J'\left( \hat{b}_{j}^{\dagger} \hat{b}_{j+2} + \textrm{H.c.} \right) 
 + V' \left( \hat{n}_j-\dfrac{1}{2}\right)\left( \hat{n}_{j+2}-\dfrac{1}{2}\right)
\right],
\end{split}
\label{rigol_hamilt}
\end{equation}
where $J$ ($J'$) is the nearest (next-nearest) neighbor hopping, 
$V$ ($V'$) is the nearest (next-nearest) neighbor interaction, and 
$V^{\textrm{ext}}_j$ is an external potential. The HCB creation 
(annihilation) operator in each site is denoted by $\hat{b}^{\dagger}_{j}$ ($\hat{b}_{j}$),  
the site number occupation by $\hat{n}_{j}=\hat{b}^{\dagger}_{j} \hat{b}_{j}$, and
$\hat{b}_{L+1}\equiv \hat{b}_{1}$ and $\hat{b}_{L+2}\equiv \hat{b}_{2}$ for periodic systems. 
Since HCBs are bosons for which the on-site repulsion is infinite, in addition to the standard commutation 
relations for bosons, their creation and annihilation operators satisfy the constraints 
$\hat{b}_{j}^{\dagger2}=\hat{b}_{j}^{2}=0$, which preclude multiple occupancy of the lattice sites.

For $J'=V'=0$, and any value of $V$, this model is integrable \cite{lieb_shultz_61}.
The approaches used to study this model are described below.

\subsection{Integrable Case with $V=J'=V'=0$ \label{rigol_intcase}}
This problem can be exactly solved if one realizes that the HCB Hamiltonian 
can be mapped onto a spin Hamiltonian by means of the Holstein--Primakoff transformation
\cite{holstein_primakoff_40},
\begin{equation}
\label{rigol_hp}
 \sigma_j^+=\hat{b}^{\dagger}_{j}\ \sqrt{1-\hat{b}^{\dagger}_{j}\hat{b}_{j}}, \quad
 \sigma_j^-=\sqrt{1-\hat{b}^{\dagger}_{j}\hat{b}_{j}}\ \hat{b}_{j}, \quad
 \sigma_j^z=\hat{b}^{\dagger}_{j}\hat{b}_{j}-1/2,
\end{equation}
and that the spin Hamiltonian can be mapped onto a noninteracting fermion Hamiltonian
utilizing the Jordan--Wigner transformation \cite{jordan_wigner_28,lieb_shultz_61} 
\begin{equation}
\label{rigol_jw}
  \sigma_j^+ = \hat{f}^{\dagger}_j\ e^{-i \pi \sum_{k<j} \hat{f}^{\dagger}_k \hat{f}_k},\quad
  \sigma_j^- = e^{i \pi \sum_{k<j} \hat{f}^{\dagger}_k \hat{f}_k}\ \hat{f}_j, \quad
  \sigma_j^z=\hat{f}^{\dagger}_j \hat{f}_j -1/2.
\end{equation}
For simplicity, we will assume open boundary conditions. The resulting Hamiltonian 
for the noninteracting fermions reads
\begin{equation}
\label{rigol_hamiltf}
  \hat{H}_\textrm{F}=- J \sum_{j=1}^{L-1} 
 \left( \hat{f}^{\dagger}_j \hat{f}_{j+1} + \textrm{H.c.} \right)
 + \sum_{j=1}^{L} V^{\textrm{ext}}_j \hat{f}^{\dagger}_j \hat{f}_{j},
\end{equation}
and, since it is quadratic, it can be easily diagonalized.
Hence, HCBs and
noninteracting fermions share the same spectrum. The density profiles and any 
density-density correlations will also coincide in both systems. The nontrivial differences
between HCBs and noninteracting fermions are revealed by the off-diagonal 
correlations. In particular, we will be interested in the time evolution of the
equal-time one-particle correlations $\hat{\rho}_{jk}$, needed to compute the momentum distribution 
function. Once again using Eq.~\eqref{rigol_hp},  
$\rho_{jk}\equiv\langle\hat{b}_{j}^{\dagger}\hat{b}_{k}\rangle=\langle\sigma_j^+\sigma_k^-\rangle=
\langle\sigma_k^-\sigma_j^+ +\delta_{jk}(1-2\sigma_j^-\sigma_j^+)\rangle$, and we focus on how 
to compute $G_{jk}=\langle\sigma_j^-\sigma_k^+\rangle$. Using \eqref{rigol_jw}, 
$G_{jk}(t)$ can be written as \cite{rigol_muramatsu_05a,rigol_muramatsu_05b}
\begin{equation}
G_{jk}(t)=\langle \Psi_{\textrm{F}}(t)| \prod_{l=1}^{j-1}
e^{i\pi \hat{f}^{\dag}_{l}\hat{f}_{l}} \hat{f}_j \hat{f}_k^{\dag}
\prod_{m=1}^{k-1} e^{-i\pi \hat{f}_{m}^{\dag}\hat{f}_{m}}
|\Psi_{\textrm{F}}(t)\rangle, 
\label{rigol_greenD1} 
\end{equation}
where $|\Psi_{\textrm{F}}(t)\rangle=e^{-i \hat{H}_{\textrm{F}}t/\hbar}|\Psi_{\textrm{F}}^{I}\rangle$, 
$|\Psi_{\textrm{F}}^{I}\rangle=\prod^{N}_{n=1}\sum^L_{q=1}P_{q n}^{I}\hat{f}^{\dag}_{q}|0 \rangle$ 
is the initial state (a  Slater determinant), $N$ is the number of particles, and $t$ 
the time. The action of exponentials whose exponents are bilinear in fermionic 
creation and annihilation operators (such as $\hat{H}_{\textrm{F}}$) on Slater determinants 
generates new Slater determinants, so 
$|\Psi_{\textrm{F}}(t)\rangle=\prod_{n=1}^{N}\sum^L_{q=1}P_{q n}(t)\hat{f}_{q}^{\dag}|0 \rangle$.
The matrix $\mathbf{P}(t)$ can be computed as 
$\mathbf{P}(t)=e^{-i\mathbf{H}_{\textrm{F}}t/\hbar}\mathbf{P}^I=\mathbf{U}e^{-i\mathbf{E}t/\hbar}\mathbf{U}^{\dagger}\mathbf{P}^I$
(where $\mathbf{H}_{\textrm{F}}$ is the corresponding matrix representation of $\hat{H}_\textrm{F})$,
and we have used that $\mathbf{H}_{\textrm{F}}\mathbf{U}=\mathbf{U}\bf{E}$, where $\bf{E}$ is a diagonal 
matrix containing the eigenenergies and $\mathbf{U}$ is the unitary matrix of 
eigenvectors. Furthermore, the action of 
$\prod_{m=1}^{k-1} e^{-i\pi \hat{f}^{\dag}_{m}\hat{f}_{m}}$ on $|\Psi_{\textrm{F}}(t)\rangle$ 
changes the sign of $P_{q n}(t)$ for $q \leq k-1$, $n=1,\ldots,N$, and the 
creation of a particle at site $k$ implies the addition of a column with only one nonzero 
element [$P_{k\,N+1}(t)=1$] [the same applies to the action of 
$\prod^{j-1}_{l=1} e^{i\pi \hat{f}^{\dag}_{l}\hat{f}_{l}} f_{j}$ on the left of 
Eq.~(\ref{rigol_greenD1})]. Hence,
\begin{align}
G_{jk}(t)
=&\langle 0 | \prod^{N}_{n=1}
\sum^L_{q=1} 
P^{*j}_{qn}(t)\hat{f}_{q} 
\prod^{N}_{l=1}
\sum^L_{m=1} 
P_{ml}^{k}(t)
\hat{f}^{\dag}_{l}\ |0 \rangle, 
\label{rigol_green1}\\
=&\det\left\{ \left[ \mathbf{P}^{j}(t)
\right]^{\dag}\mathbf{P}^{k}(t)\right\}.
\label{rigol_green2}
\end{align}
In Eq.~\eqref{rigol_green1}, the matrix elements $P^{j}_{qn}(t)$ and $P^{k}_{ml}(t)$ 
have the form
\begin{equation}
 P^{\alpha}_{\beta \gamma}(t)= \left\{ \begin{array}{rl}
 -P_{\beta \gamma}(t) & \text{for } \beta<    \alpha,\,\gamma=1,\ldots,N \\
\,P_{\beta \gamma}(t) & \text{for } \beta\geq \alpha,\,\gamma=1,\ldots,N \\
  \delta_{\alpha\beta} & \text{for } \gamma=N + 1
\end{array}\right.,
\end{equation}
with $\alpha=j,k$, $\beta=q,m$, and $\gamma=n,l$. Equation~\eqref{rigol_green2} 
follows from \eqref{rigol_green1} by using the identity
\begin{equation}
\langle 0 |\hat{f}_{\alpha_1}\cdot \cdot \cdot \hat{f}_{\alpha_{N+1}}
\hat{f}^{\dag}_{\beta_{N+1}} \cdot \cdot \cdot \hat{f}^{\dag}_{\beta_1}|0 \rangle
=\epsilon^{\lambda_1\cdot \cdot \cdot \lambda_{N+1}}
\delta_{\alpha_1\beta_{\lambda_1}}\cdot \cdot \cdot
\delta_{\alpha_{N+1}\beta_{\lambda_{N+1}}},
\end{equation}
where $\epsilon^{\lambda_1\cdot \cdot \cdot \lambda_{N+1}}$ is 
the Levi-Civita symbol in $N+1$ dimensions, and the indices $\lambda$ 
have values between 1 and $N+1$. Employing Eq.~\eqref{rigol_green2}, $\rho_{jk}$ can be 
calculated in polynomial time, scaling as $L^2N^3$, using a computer. 

We will also be interested in describing the momentum distribution function after relaxation 
by using statistical ensembles. A polynomial time approach in this case is only known to us
within the grand-canonical formalism \cite{rigol_05}. The one-particle density matrix in this ensemble can be written as
\begin{equation}
\begin{split}
\rho_{jk} \equiv& \frac{1}{Z}\textrm{Tr}\left\lbrace 
\hat{b}^{\dagger}_{j}\hat{b}_{k}e^{-(\hat{H}_{\textrm{HCB}}-\mu\sum_{n} \hat{b}^{\dagger}_{n}
\hat{b}_{n})/k_BT}\right\rbrace 
\\ = &
\dfrac{1}{Z}\textrm{Tr}\left\lbrace  \hat{f}_{j}^{\dagger}\hat{f}_{k} 
\prod^{k-1}_{l=1} e^{i\pi \hat{f}_{l}^{\dagger}\hat{f}_{l}}e^{-(\hat{H}_{\textrm{F}}-\mu\sum_{n} 
\hat{f}^{\dagger}_{n}\hat{f}_{n})/k_B T} 
\prod^{j-1}_{l=1} e^{-i\pi \hat{f}^{\dagger}_{m}\hat{f}_{m}}  \right\rbrace ,
\end{split}
\label{rigol_grand1}
\end{equation}
where $\mu$ is the chemical potential, $k_B$ is Boltzmann's constant, $T$ the temperature, 
and $Z=\textrm{Tr} \lbrace 
e^{-(\hat{H}_{\textrm{HCB}} -\mu\sum_{n} \hat{b}^{\dagger}_{n}\hat{b}_{n})/k_B T}\rbrace$ (identical for HCBs and fermions) is the partition function. To arrive to 
Eq.~\eqref{rigol_grand1}, in addition to the Jordan--Wigner transformation [Eq.~\eqref{rigol_jw}], we have used the 
cyclic property of the trace. Another useful property of the trace, over the fermionic Fock space \cite{rigol_05}, is 
\begin{equation}\label{rigol_tr}
\textrm{Tr}\left\lbrace
e^{\sum_{jk}\hat{f}^{\dagger}_{j} X_{jk}\hat{f}_{k}} 
e^{\sum_{lm}\hat{f}^{\dagger}_{l} Y_{lm}\hat{f}_{m}} \cdots 
e^{\sum_{nq}\hat{f}^{\dagger}_{n} Z_{nq}\hat{f}_{q}} \right\rbrace
=\det\left[\mathbf{I}+ e^{\mathbf{X}}e^{\mathbf{Y}}\cdots e^{\mathbf{Z}}\right], 
\end{equation}
where $\mathbf{I}$ is the identity matrix. Equation \eqref{rigol_tr} allows us to compute $Z$ as
$Z=\prod_j \left[ 1+e^{-(E_{jj}-\mu)/k_B T} \right]$. By noticing that for $j\neq k$, we can write 
$f^{\dagger}_{j}f_{k}= \exp\left( \sum_{nq}f^{\dagger}_{n} A_{nq}f_{q}\right) -1$,
where the only nonzero element of $\mathbf{A}$ is $A_{jk}=1$, the off-diagonal elements of 
$\rho_{jk}$ ($j\neq k$) can be obtained as
\begin{equation}
\begin{split}
\rho_{jk} =& \dfrac{1}{Z}\left\lbrace  \det \left[\mathbf{I}+ 
(\mathbf{I}+\mathbf{A})\mathbf{O}_{1} \mathbf{U} e^{-(\mathbf{E}-\mu \mathbf{I})/k_B T} 
\mathbf{U}^{\dagger} \mathbf{O}_{2}\right]
\right. \\ & \left.  
-\det\left[\mathbf{I}+ \mathbf{O}_{1} \mathbf{U} 
e^{-(\mathbf{E}-\mu \mathbf{I})/k_B T}\mathbf{U}^{\dagger} \mathbf{O}_{2}\right]
\right\rbrace,
\end{split}
\end{equation}
where $\mathbf{O}_{1}$ ($\mathbf{O}_{2}$) is diagonal with the first $j-1$ ($k-1$) 
elements of the diagonal equal to $-1$ and the others equal to $1$.
The diagonal elements of 
$\rho_{jk}$ are the same as for noninteracting fermions and can be computed as 
\begin{equation}
\rho_{jj}=\left[\mathbf{I}+ 
e^{-(\mathbf{H}_{\textrm{F}}-\mu \mathbf{I})/k_B T}\right]^{-1}_{jj}  =
\left[\mathbf{U} \left(\mathbf{I}+e^{-(\mathbf{E}-\mu \mathbf{I})/k_B T}\right)^{-1}  
\mathbf{U}^{\dagger} \right]_{jj}.
\end{equation}
The computational time within this approach scales as $L^5$. 

The momentum distribution function $n(k)$ in and out of equilibrium, is then determined by the expression $n(k)=(1/L) \sum_{mn} e^{-i k(m-n)} \rho_{nm}$.

\subsection{Nonintegrable Case with $V^{\rm ext}=0$}
For this case, we make use of full exact diagonalization (see, e.g., \cite{numerical_recipes_07}).  This approach has the disadvantage 
that the dimension of the matrices needing to be diagonalized scales exponentially with system size. Since the Hamiltonian 
\eqref{rigol_hamilt} conserves the total number of particles, we work with a fixed number of particles $N=L/3$, reducing the 
dimensionality of our problem from $2^L$ to $\binom{L}{N}$.
To further reduce the dimensionality of the matrices to be diagonalized, we consider systems with periodic boundary conditions 
and no external potential ($V^{\textrm{ext}}=0$). Then, by using translational symmetry, we can block-diagonalize the Hamiltonian, 
with the size of each momentum block being $\sim 1/L$ the size of the original matrix. All momentum sectors, the dimensions of 
which are shown in the table below \cite{santos_rigol_10STATc},
are diagonalized. They are all used to construct the microcanonical 
and canonical ensembles.
\begin{center}
\begin{tabular}{|c||c|c|c|c|}
\hline
\multicolumn{5}{|c|}{Dimension of all momentum sectors ($k=2\pi \kappa/L$)}\\
\hline
\hline
$L=18$ & $\kappa=0,6 $ & $\kappa=1,5,7 $ & $\kappa=2,4,8 $ & $\kappa=3,9$ \\
\hline
dimension & 1038 & 1026 & 1035 & 1028 \\
\hline
\hline
$L=21$ & $\kappa=0,7$ & other $\kappa$'s & &  \\
\hline
dimension  & 5538 & 5537 &  & \\
\hline
\hline
$L=24$ & $\kappa=0,8$ & $\kappa=4,12$ & $\kappa=2,6,10$ & odd $\kappa$'s  \\
\hline
dimension & 30667 & 30666 & 30664 & 30624 \\
\hline
\end{tabular}
\end{center}

\section{Results}
We focus on the dynamics after a sudden quench. This means that we start with some 
eigenstate of an initial Hamiltonian, which may not be the 
ground state, then at $t=0$ some parameter is changed and the 
system is allowed to evolve. Independently of whether the Hamiltonian is integrable or not, 
one can always write the initial state wavefunction $\vert\psi_{\textrm{ini}}\rangle$
in the eigenstate basis of the final Hamiltonian, i.e., 
$\vert\psi_{\textrm{ini}}\rangle=\sum_\alpha C_{\alpha}\vert\Psi_\alpha\rangle$
with $C_{\alpha}=\langle\Psi_\alpha\vert\psi_{\textrm{ini}}\rangle$ and 
$\hat{H}\vert\Psi_\alpha\rangle=E_\alpha\vert\Psi_\alpha\rangle$. The dynamics of the 
wave-function takes the form $\vert\psi(t)\rangle=e^{-i\hat{H}t/\hbar}\vert\psi_{\textrm{ini}}\rangle
=\sum_\alpha e^{-iE_\alpha t/\hbar} C_{\alpha}\vert\Psi_\alpha\rangle$ and the expectation
value of any observable $\hat{O}$ can be written as $\langle \hat{O}(t)\rangle\equiv
\langle \psi(t) | \hat{O} | \psi(t)  \rangle =\sum_{\alpha,\beta} C_{\alpha}^{*} C_{\beta}
e^{i(E_{\alpha}-E_{\beta})t} O_{\alpha\beta}$, where 
$O_{\alpha\beta}=\langle\Psi_\alpha\vert\hat{O}\vert\Psi_\beta\rangle$. If the spectrum is 
nondegenerate, the infinite time average and the observable after relaxation
is determined by
\begin{equation}
\overline{\langle \hat{O} \rangle} \equiv O_{\textrm{diag}}=\sum_{\alpha} |C_{\alpha}|^{2}
O_{\alpha\alpha}.
\label{rigol_diagonal}
\end{equation}
This exact result can be thought as the prediction of a `diagonal ensemble,' 
where $|C_{\alpha}|^{2}$ is the weight of each state \cite{rigol_dunjko_08}, 
and is different from any conventional ensemble of  statistical mechanics.

\subsection{Integrable Case with $V=J'=V'=0$}
Here, our set up is close in spirit to that of the experiment 
\cite{kinoshita_wenger_06}. The initial state is the ground state of a harmonic trap with a 
staggered potential and, at $t=0$, we turn off the staggered potential and allow the 
system to evolve in the presence of the trap \cite{rigol_muramatsu_06}. In addition to density 
profiles and $n(k)$, we also study the occupation of the natural orbitals, which are the 
eigenstates of the one-particle density matrix, determined by the eigenvalue equation 
$\sum^N_{k=1} \rho_{jk}\phi_{k}^{\eta}=\lambda_{\eta}\phi^\eta_{j}$. The lowest natural orbital 
is also the most highly occupied.

\begin{figure}[!t]
\centerline{\includegraphics[width=0.83\textwidth]{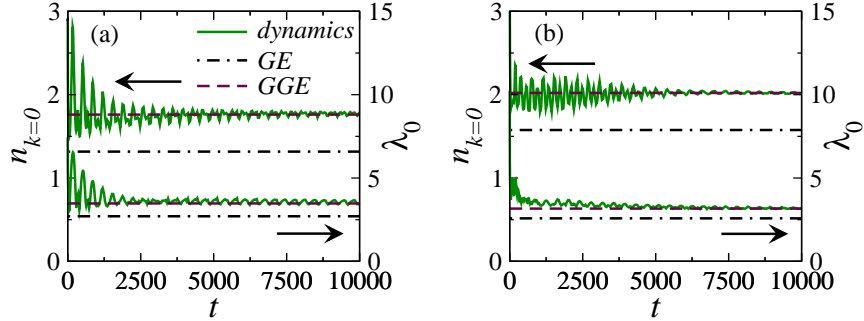}}
\caption{Dynamics of $n(k=0)$ (top plots) and $\lambda_0$ (bottom plots) 
after a staggered potential is turned off in harmonically confined systems with 900 lattice 
sites and a trap curvature $V_2=3\times 10^{-5}J$. $t$ is given in units of $\hbar/J$, and 
the evolution starts from the ground state in the presence of a staggered potential of 
strength $0.5J$. The number of particles is: (a) $N=200$ 
and (b) $N=299$. [The corresponding initial density profiles can be seen in 
Figs.\ \ref{rigol_TrapLattOff_Perfiles}(a) and \ref{rigol_TrapLattOff_Perfiles}(d)]. 
The dashed-dotted lines depict the results within the grand-canonical ensemble (GE), with
(a) $T=0.31J$ and (b) $T=0.33J$, and the dashed lines the results within the generalized 
Gibbs ensemble (GGE).
}
\label{rigol_TrapLattOff_nk0no0vstau}
\end{figure}

\begin{figure}[!t]
\centerline{\includegraphics[width=0.99\textwidth]{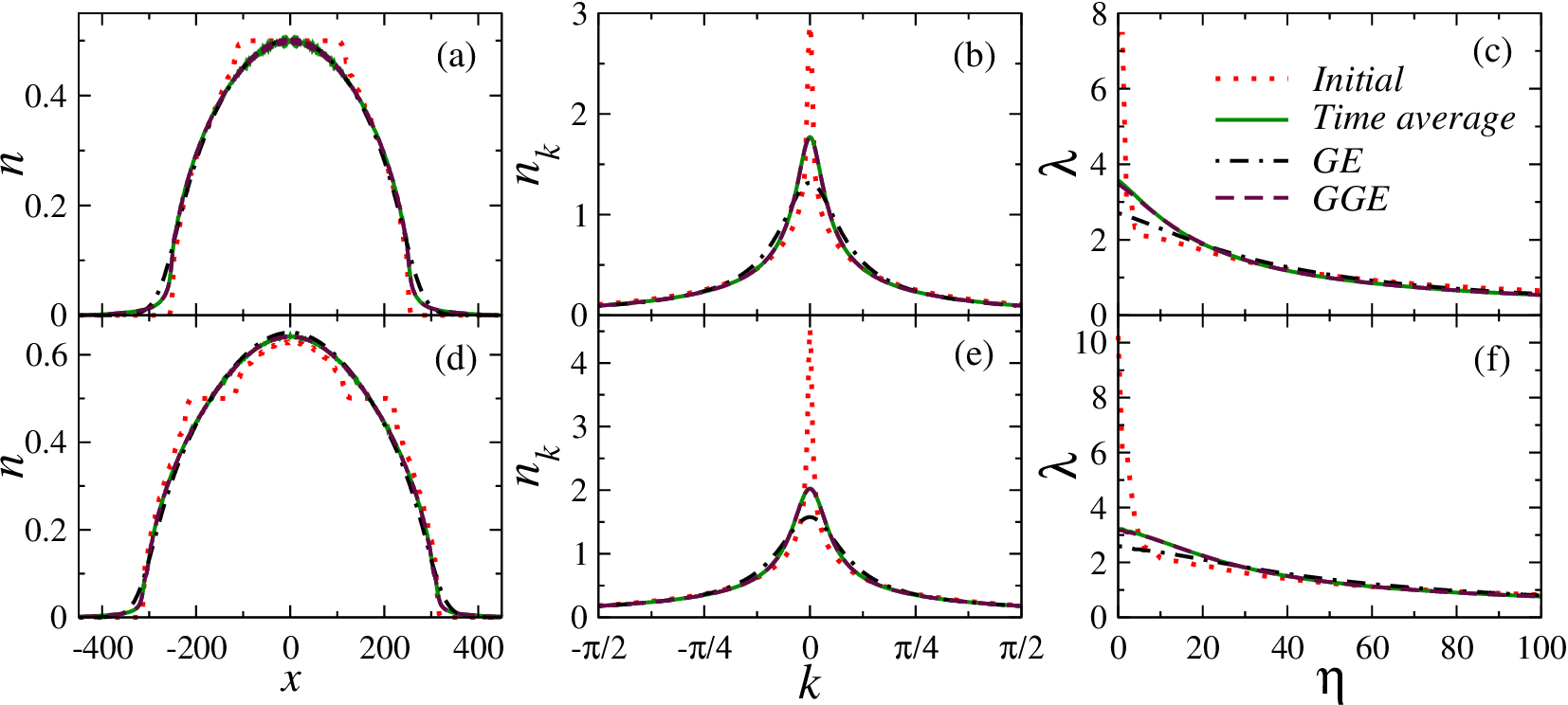}}
\caption{Initial state and time average values of: (a),(d) density profiles, (b),(e) momentum 
distribution functions, and (c),(f) occupation of the lowest 100 natural orbitals. The averages 
are computed between $t=5000\hbar/J$ and $t=10000\hbar/J$ with measurements done in time intervals 
$\Delta t=40 \hbar/J$, and correspond to the dynamics depicted in 
Fig.\ \ref{rigol_TrapLattOff_nk0no0vstau}. The results of the time average are compared with 
those obtained in the grand-canonical ensemble (GE) and the generalized Gibbs ensemble (GGE) described in the text. The number of particles is $N=200$ (a)--(c) and $N=299$ (d)--(f). 
In (a) and (d), for the initial state, the occupations plotted are the averaged density per unit cell. 
Note that in the presence of the staggered potential, the density exhibits large fluctuations 
from site to site. Flat regions of the unit cell occupations correspond to insulating domains \cite{rigol_muramatsu_06}.}
\label{rigol_TrapLattOff_Perfiles}
\end{figure}

Figure \ref{rigol_TrapLattOff_nk0no0vstau} depicts the evolution of
the occupation of the zero-momentum state $n(k=0)$ and the lowest 
natural orbital $\lambda_0$ when, (i) the initial state has a half-filled
insulator in the center of the trap [Fig.\ \ref{rigol_TrapLattOff_Perfiles}(a)]
and, (ii) two insulating shoulders surround a central superfluid region
[Fig.\ \ref{rigol_TrapLattOff_Perfiles}(d)]. In both cases, the two observables 
undergo relaxation dynamics, which ultimately brings them to an almost time-independent result. 
This shows that relaxation is not precluded by integrability,
and the question that remains to be answered is how to describe these observables 
after relaxation. As seen in Fig.~\ref{rigol_TrapLattOff_nk0no0vstau}, they are clearly 
different from the predictions of the grand-canonical ensemble (GE in the figures), 
which are obtained after determining the temperature and chemical potential so that 
\begin{equation}
E=\frac{1}{Z}\textrm{Tr}\left\lbrace\hat{H}e^{-(\hat{H}-\mu\hat{N})/k_BT}\right\rbrace,
\quad N=\frac{1}{Z}\textrm{Tr}\left\lbrace\hat{N}e^{-(\hat{H}-\mu\hat{N})/k_BT}\right\rbrace,\quad 
\end{equation}
where $\hat{N}=\sum_{j}\hat{b}^{\dagger}_{j}\hat{b}_{j}$, and $E$ and $N$ are the average energy
and particle number in the time evolving state, which are conserved during the evolution.
We note that for the system sizes considered, finite size effects are negligible.

The lack of relaxation to the thermal state may not be surprising considering that the
system is integrable, and, hence, the existence of conserved quantities may preclude
thermalization. In Ref.\ \cite{rigol_dunjko_07}, a generalization of the Gibbs ensemble 
(GGE) was proposed in order to account for the conserved quantities and still be able
to statistically describe integrable systems. The density matrix for the GGE was determined
by maximizing the many-body Gibbs entropy $S=k_B \textrm{Tr}\left[\hat{\rho}_c 
\ln(1/\hat{\rho}_c) \right]$ subject to the constraints imposed by all 
the integrals of motion. The result reads
\begin{equation}
\hat{\rho}_c=\frac{1}{Z_c}e^{-\sum_{j=1}^{L} \lambda_{j} \hat{I}_{j}}, \quad
Z_c=\textrm{Tr}\left\lbrace e^{-\sum_{j=1}^{L} \lambda_{j} \hat{I}_{j}}\right\rbrace
\label{rigol_constrained}
\end{equation}
where $Z_c$ is the generalized partition function, $\lbrace \hat{I}_{j} \rbrace$ is
a full set of integrals of motion,  and $\lbrace \lambda_{j} \rbrace$ are 
the Lagrange multipliers. The Lagrange multipliers are computed 
using the expectation values of the full set of integrals of motion in the initial state, i.e.,
$\langle\hat{I}_{j}\rangle_{\textrm{ini}}=\textrm{Tr}\lbrace\hat{I}_{j}\hat{\rho}_c\rbrace$.
For HCBs, which can be mapped to noninteracting fermions, a natural set of integrals
of motion is provided by the projection operators to the noninteracting single particle
eigenstates $\lbrace\hat{I}_{j}\rbrace=\lbrace\hat{\gamma}^{f\dagger}_{j}\hat{\gamma}^{f}_{j}\rbrace$,
where $\lbrace\hat{\gamma}^{f\dagger}_{j}\rbrace$ ($\lbrace\hat{\gamma}^{f}_{j}\rbrace$) creates
(annihilates) a single particle in an eigenstate of Eq.~\eqref{rigol_hamiltf}. The resulting
Lagrange multipliers read $\lambda_{j}=\ln[(1-\langle\hat{I}_{j}\rangle_{\textrm{ini}})/
\langle\hat{I}_{j}\rangle_{\textrm{ini}}]$. They allow one to build the density matrix in 
Eq.~\eqref{rigol_constrained} and to compute expectation values as was described for the 
grand-canonical ensemble in Sec.~\ref{rigol_intcase}.

Figure \ref{rigol_TrapLattOff_nk0no0vstau} shows that the GGE calculations for 
$n(k=0)$ and $\lambda_0$ properly predict the outcome of the relaxation dynamics. 
We have also computed the time average (between $t=5000\hbar/J$ and $10000\hbar/J$) 
of the full density profiles, $n(k)$, and $\lambda_{\eta}$. They are shown in 
Fig.~\ref{rigol_TrapLattOff_Perfiles}. There, the time averages are compared with 
the results for the initial state and with the predictions of the GE and the GGE. That
comparison clearly shows that, unlike the GE, the GGE is able to predict all those 
single particle observables after relaxation. Note that, when  
written in the bosonic language, the constraints lose the bilinear character 
they have in the fermionic representation, i.e., the outcome of the GGE calculation
is not at all trivial, as it would be if done for noninteracting fermions. Recent
numerical and analytical studies have addressed various aspects of the GGE
\cite{rigol_dunjko_07,cazalilla_06,rigol_muramatsu_06,calabrese_cardy_07,cramer_dawson_08,
barthel_schollwock_08,kollar_eckstein_08,iucci_cazalilla_09}, while a microscopic understanding 
for the agreement between the predictions of the GGE and the diagonal ensemble was presented 
in Ref.~\cite{cassidy_clark_11}.

\subsection{Nonintegrable Case with $V^{\rm ext}=0$}
To study the effects of breaking integrability, we prepare an initial state that is an eigenstate 
of a Hamiltonian (in the total momentum $k=0$ sector) with $J_{\textrm{ini}}$, $V_{\textrm{ini}}$, $J'$, $V'$ 
and then quench the nearest-neighbor parameters to $J_{\textrm{fin}}$, $V_{\textrm{fin}}$ without changing $J'$, 
$V'$. The same quench is repeated for different values of $J',V'$ as one departs from $J'=V'=0$ 
\cite{rigol_09a}. To find whether the dynamics brings the observables to the predictions of the 
diagonal ensemble \eqref{rigol_diagonal}, we calculate the normalized area between the 
observables during the time evolution and their infinite time average, i.e., 
$\delta n_k(t)=(\sum_k |n(k,t)-n_{\textrm{diag}}(k)|)/\sum_k n_{\textrm{diag}}(k)$. Similarly, we compute
$\delta N_k$ for the structure factor $N(k)$, which is the Fourier transform of the density-density 
correlations. 

\begin{figure}[!b]
\centerline{\includegraphics[width=0.92\textwidth,angle=0]{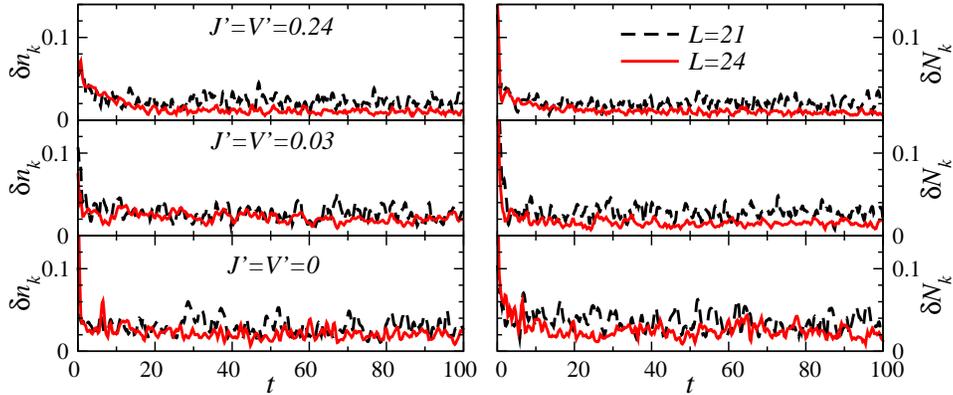}}
\caption{\label{rigol_TimeEvolution_L24T2.0}
Evolution of $\delta n_k$ (left panels) and $\delta N_k$ (right panels) after a quench
from $J_{\textrm{ini}}=0.5J$, $V_{\textrm{ini}}=2.0J$ to $J_{\textrm{fin}}=J$, $V_{\textrm{fin}}=J$, with $J'_{ini}=J'_{fin}=J'$ 
and $V'_{ini}=V'_{fin}=V'$, for two system sizes. The initial state was selected within the 
eigenstates with total momentum $k=0$ such that after the quench the effective temperature is 
$T=3.0$ in all cases. Given the energy of the initial state $E$, $T$ follows from 
$E=Z^{-1}\textrm{Tr}\lbrace\hat{H}e^{-\hat{H}/k_B T}\rbrace$, 
where $Z=\textrm{Tr}\lbrace e^{-\hat{H}/k_BT}\rbrace$. 
The trace runs over the full spectrum.
}
\end{figure}

In Fig.\ \ref{rigol_TimeEvolution_L24T2.0}, we show results for $\delta n_k$ and $\delta N_k$ 
vs $t$ for three different quenches and two system sizes. The time evolution 
is very similar in all cases, and is consistent with a fast relaxation of both observables
towards the diagonal ensemble prediction (in a time scale $t\sim \hbar/J$). The 
average differences after relaxation and their fluctuations can be seen to decrease
with increasing system size. From these results, we infer that, for very large systems 
sizes, $n(k)$ and $N(k)$ should in general relax to exactly the predictions of 
Eq.~\eqref{rigol_diagonal} even if the system is very close or at integrability. 

\begin{figure}[!t]
\centerline{\includegraphics[width=0.9\textwidth,angle=0]{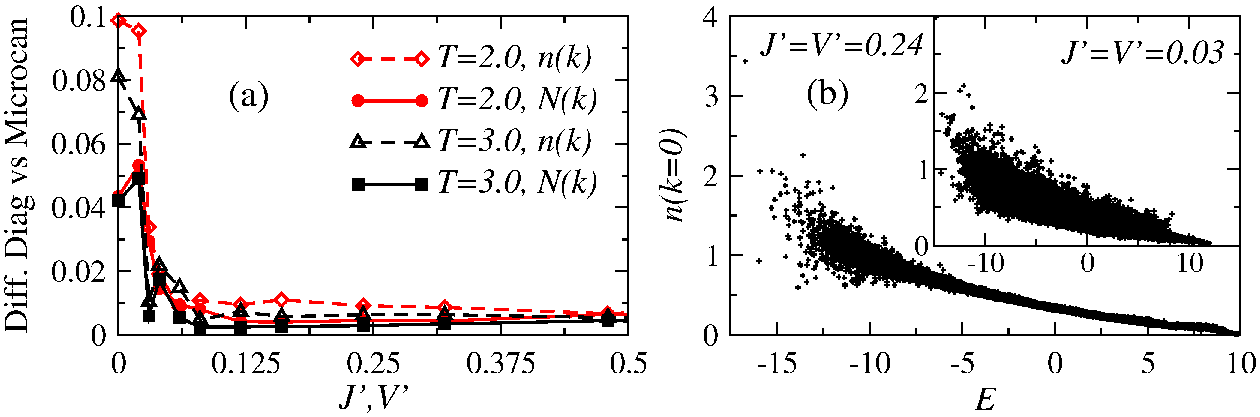}}
\caption{\label{rigol_ETH}
(a) Differences between the predictions of the diagonal and microcanonical ensembles 
(calculated as $\delta n_k$ and $\delta N_k$ in Fig.~\ref{rigol_TimeEvolution_L24T2.0}).
Results are shown for $T=2.0$ and $T=3.0$. (b) $n(k=0)$ as a function of the energy for all 
the eigenstates of the Hamiltonian (including all momentum sectors). 
(main panel) $J=V=1$ and $J'=V'=0.24$. (inset) $J=V=1$ and $J'=V'=0.03$. 
The systems in (a) and (b) have $L=24$ and $N_b=8$.
}
\end{figure}

We then say that thermalization takes place if the results of conventional statistical 
ensembles and those of the diagonal ensemble are the same.
In Fig.\ \ref{rigol_ETH}(a), we compare the diagonal ensemble results with the 
predictions of the microcanonical ensemble for our two observables of interest. Far 
from integrability the differences are small and decrease with increasing
system size \cite{rigol_09a}, i.e., thermalization takes place. As one approaches 
integrability, the differences increase, signaling a breakdown of thermalization in 1d.

Thermalization away from integrability, as well as its failure close to integrability,
can be understood in terms of the eigenstate thermalization hypothesis (ETH)
\cite{rigol_dunjko_08,deutsch_91,srednicki_94}. ETH states that, for generic systems, 
the fluctuations of eigenstate expectation values of observables is small between eigenstates 
that are close in energy, which implies that the microcanonical average is identical 
to the prediction of each eigenstate, which is the same as saying that the eigenstates already 
exhibit thermal behavior. If this holds, thermalization in an isolated quantum system will 
follow for any distribution of $|C_{\alpha}|^{2}$ that is narrow enough in energy. 

The main panel in Fig.\ \ref{rigol_ETH}(b) depicts $n(k=0)$ [similar results were obtained 
for $n(k\neq0)$ and $N(k)$] in each eigenstate of the Hamiltonian when the system
is far from integrability. After a region of low energies where the eigenstate expectation 
values exhibit large fluctuations, one can see another region where fluctuations are small
(presumably vanishing in the thermodynamic limit)
and ETH holds. The inset shows that for a system close to the integrable point, in which 
thermalization is absent [Fig.\ \ref{rigol_ETH}(a)], the eigenstate-to-eigenstate 
fluctuations of $n(k=0)$ are very large over the entire spectrum (they do not vanish
in the thermodynamic limit) and ETH does not hold.

\section*{Acknowledgments}
This work was supported by the Office of Naval Research. We thank V.~Dunjko, A.~Muramatsu, 
M.~Olshanii, and L.~F.~Santos for discussions.


\begin{thebibliography}{46}%
\makeatletter
\providecommand \@ifxundefined [1]{%
 \@ifx{#1\undefined}
}%
\providecommand \@ifnum [1]{%
 \ifnum #1\expandafter \@firstoftwo
 \else \expandafter \@secondoftwo
 \fi
}%
\providecommand \@ifx [1]{%
 \ifx #1\expandafter \@firstoftwo
 \else \expandafter \@secondoftwo
 \fi
}%
\providecommand \natexlab [1]{#1}%
\providecommand \enquote  [1]{``#1''}%
\providecommand \bibnamefont  [1]{#1}%
\providecommand \bibfnamefont [1]{#1}%
\providecommand \citenamefont [1]{#1}%
\providecommand \href@noop [0]{\@secondoftwo}%
\providecommand \href [0]{\begingroup \@sanitize@url \@href}%
\providecommand \@href[1]{\@@startlink{#1}\@@href}%
\providecommand \@@href[1]{\endgroup#1\@@endlink}%
\providecommand \@sanitize@url [0]{\catcode `\\12\catcode `\$12\catcode
  `\&12\catcode `\#12\catcode `\^12\catcode `\_12\catcode `\%12\relax}%
\providecommand \@@startlink[1]{}%
\providecommand \@@endlink[0]{}%
\providecommand \url  [0]{\begingroup\@sanitize@url \@url }%
\providecommand \@url [1]{\endgroup\@href {#1}{\urlprefix }}%
\providecommand \urlprefix  [0]{URL }%
\providecommand \Eprint [0]{\href }%
\providecommand \doibase [0]{http://dx.doi.org/}%
\providecommand \selectlanguage [0]{\@gobble}%
\providecommand \bibinfo  [0]{\@secondoftwo}%
\providecommand \bibfield  [0]{\@secondoftwo}%
\providecommand \translation [1]{[#1]}%
\providecommand \BibitemOpen [0]{}%
\providecommand \bibitemStop [0]{}%
\providecommand \bibitemNoStop [0]{.\EOS\space}%
\providecommand \EOS [0]{\spacefactor3000\relax}%
\providecommand \BibitemShut  [1]{\csname bibitem#1\endcsname}%
\let\auto@bib@innerbib\@empty
\bibitem{gallavotti_book_99}%
  \BibitemOpen
  \bibfield  {author} {\bibinfo {author} {\bibfnamefont {G.}~\bibnamefont
  {Gallavotti}},\ }\href@noop {} {\emph {\bibinfo {title} {Statistical
  Mechanics: A Short Treatise}}}\ (\bibinfo  {publisher} {Springer},\ \bibinfo
  {address} {Berlin},\ \bibinfo {year} {1999})\BibitemShut {NoStop}%
\bibitem{fermi_pasta_55}%
  \BibitemOpen
  \bibfield  {author} {\bibinfo {author} {\bibfnamefont {E.}~\bibnamefont
  {Fermi}}, \bibinfo {author} {\bibfnamefont {J.}~\bibnamefont {Pasta}}, and\
  \bibinfo {author} {\bibfnamefont {S.}~\bibnamefont {Ulam}},\ }\href@noop {}
  {\bibfield  {journal} {\bibinfo  {journal} {Los Alamos Report}\ ,\ \bibinfo
  {pages} {LA}} (\bibinfo {year} {1955})}\BibitemShut {NoStop}%
\bibitem{campbell_rosenau_05}%
  \BibitemOpen
  \bibfield  {author} {\bibinfo {author} {\bibfnamefont {D.~K.}\ \bibnamefont
  {Campbell}}, \bibinfo {author} {\bibfnamefont {P.}~\bibnamefont {Rosenau}}, 
  and\ \bibinfo {author} {\bibfnamefont {G.~M.}\ \bibnamefont {Zaslavsky}},\
  }\href@noop {} {\bibfield  {journal} {\bibinfo  {journal} {Chaos}\ }\textbf
  {\bibinfo {volume} {15}},\ \bibinfo {pages} {015101} (\bibinfo {year}
  {2005})}\BibitemShut {NoStop}%
\bibitem{rasetti_book_86}%
  \BibitemOpen
  \bibfield  {author} {\bibinfo {author} {\bibfnamefont {M.}~\bibnamefont
  {Rasetti}},\ }\href@noop {} {\emph {\bibinfo {title} {Modern Methods in
  Equilibrium Statistical Mechanics}}}\ (\bibinfo  {publisher} {World
  Scientific},\ \bibinfo {address} {Singapore},\ \bibinfo {year}
  {1986})\BibitemShut {NoStop}%
\bibitem{kinoshita_wenger_04}%
  \BibitemOpen
  \bibfield  {author} {\bibinfo {author} {\bibfnamefont {T.}~\bibnamefont
  {Kinoshita}}, \bibinfo {author} {\bibfnamefont {T.}~\bibnamefont {Wenger}}, 
  and\ \bibinfo {author} {\bibfnamefont {D.~S.}\ \bibnamefont {Weiss}},\
  }\href@noop {} {\bibfield  {journal} {\bibinfo  {journal} {Science}\ }\textbf
  {\bibinfo {volume} {305}},\ \bibinfo {pages} {1125} (\bibinfo {year}
  {2004})}\BibitemShut {NoStop}%
\bibitem{paredes_widera_04}%
  \BibitemOpen
  \bibfield  {author} {\bibinfo {author} {\bibfnamefont {B.}~\bibnamefont
  {Paredes}}, \bibinfo {author} {\bibfnamefont {A.}~\bibnamefont {Widera}},
  \bibinfo {author} {\bibfnamefont {V.}~\bibnamefont {Murg}}, \bibinfo {author}
  {\bibfnamefont {O.}~\bibnamefont {Mandel}}, \bibinfo {author} {\bibfnamefont
  {S.}~\bibnamefont {F\"olling}}, \bibinfo {author} {\bibfnamefont
  {I.}~\bibnamefont {Cirac}}, \bibinfo {author} {\bibfnamefont {G.~V.}\
  \bibnamefont {Shlyapnikov}}, \bibinfo {author} {\bibfnamefont {T.~W.}\
  \bibnamefont {H\"ansch}},  and\ \bibinfo {author} {\bibfnamefont
  {I.}~\bibnamefont {Bloch}},\ }\href@noop {} {\bibfield  {journal} {\bibinfo
  {journal} {Nature}\ }\textbf {\bibinfo {volume} {429}},\ \bibinfo {pages}
  {277} (\bibinfo {year} {2004})}\BibitemShut {NoStop}%
\bibitem{kinoshita_wenger_05}%
  \BibitemOpen
  \bibfield  {author} {\bibinfo {author} {\bibfnamefont {T.}~\bibnamefont
  {Kinoshita}}, \bibinfo {author} {\bibfnamefont {T.}~\bibnamefont {Wenger}}, 
  and\ \bibinfo {author} {\bibfnamefont {D.~S.}\ \bibnamefont {Weiss}},\
  }\href@noop {} {\bibfield  {journal} {\bibinfo  {journal} {Phys. Rev. Lett.}\
  }\textbf {\bibinfo {volume} {95}},\ \bibinfo {pages} {190406} (\bibinfo
  {year} {2005})}\BibitemShut {NoStop}%
\bibitem{kinoshita_wenger_06}%
  \BibitemOpen
  \bibfield  {author} {\bibinfo {author} {\bibfnamefont {T.}~\bibnamefont
  {Kinoshita}}, \bibinfo {author} {\bibfnamefont {T.}~\bibnamefont {Wenger}}, 
  and\ \bibinfo {author} {\bibfnamefont {D.~S.}\ \bibnamefont {Weiss}},\
  }\href@noop {} {\bibfield  {journal} {\bibinfo  {journal} {Nature}\ }\textbf
  {\bibinfo {volume} {440}},\ \bibinfo {pages} {900} (\bibinfo {year}
  {2006})}\BibitemShut {NoStop}%
\bibitem{hofferberth_lesanovsky_07}%
  \BibitemOpen
  \bibfield  {author} {\bibinfo {author} {\bibfnamefont {S.}~\bibnamefont
  {Hofferberth}}, \bibinfo {author} {\bibfnamefont {I.}~\bibnamefont
  {Lesanovsky}}, \bibinfo {author} {\bibfnamefont {B.}~\bibnamefont {Fischer}},
  \bibinfo {author} {\bibfnamefont {T.}~\bibnamefont {Schumm}},  and\ \bibinfo
  {author} {\bibfnamefont {J.}~\bibnamefont {Schmiedmayer}},\ }\href@noop {}
  {\bibfield  {journal} {\bibinfo  {journal} {Nature}\ }\textbf {\bibinfo
  {volume} {449}},\ \bibinfo {pages} {324} (\bibinfo {year}
  {2007})}\BibitemShut {NoStop}%
\bibitem{rigol_dunjko_07}%
  \BibitemOpen
  \bibfield  {author} {\bibinfo {author} {\bibfnamefont {M.}~\bibnamefont
  {Rigol}}, \bibinfo {author} {\bibfnamefont {V.}~\bibnamefont {Dunjko}},
  \bibinfo {author} {\bibfnamefont {V.}~\bibnamefont {Yurovsky}},  and\
  \bibinfo {author} {\bibfnamefont {M.}~\bibnamefont {Olshanii}},\ }\href@noop
  {} {\bibfield  {journal} {\bibinfo  {journal} {Phys. Rev. Lett.}\ }\textbf
  {\bibinfo {volume} {98}},\ \bibinfo {pages} {050405} (\bibinfo {year}
  {2007})}\BibitemShut {NoStop}%
\bibitem{cazalilla_06}%
  \BibitemOpen
  \bibfield  {author} {\bibinfo {author} {\bibfnamefont {M.~A.}\ \bibnamefont
  {Cazalilla}},\ }\href@noop {} {\bibfield  {journal} {\bibinfo  {journal}
  {Phys. Rev. Lett.}\ }\textbf {\bibinfo {volume} {97}},\ \bibinfo {pages}
  {156403} (\bibinfo {year} {2006})}\BibitemShut {NoStop}%
\bibitem{rigol_muramatsu_06}%
  \BibitemOpen
  \bibfield  {author} {\bibinfo {author} {\bibfnamefont {M.}~\bibnamefont
  {Rigol}}, \bibinfo {author} {\bibfnamefont {A.}~\bibnamefont {Muramatsu}}, 
  and\ \bibinfo {author} {\bibfnamefont {M.}~\bibnamefont {Olshanii}},\
  }\href@noop {} {\bibfield  {journal} {\bibinfo  {journal} {Phys. Rev. A}\
  }\textbf {\bibinfo {volume} {74}},\ \bibinfo {pages} {053616} (\bibinfo
  {year} {2006})}\BibitemShut {NoStop}%
\bibitem{calabrese_cardy_07}%
  \BibitemOpen
  \bibfield  {author} {\bibinfo {author} {\bibfnamefont {P.}~\bibnamefont
  {Calabrese}} and\ \bibinfo {author} {\bibfnamefont {J.}~\bibnamefont
  {Cardy}},\ }\href@noop {} {\bibfield  {journal} {\bibinfo  {journal} {J.
  Stat. Mech.}\ ,\ \bibinfo {pages} {P06008}} (\bibinfo {year}
  {2007})}\BibitemShut {NoStop}%
\bibitem{cramer_dawson_08}%
  \BibitemOpen
  \bibfield  {author} {\bibinfo {author} {\bibfnamefont {M.}~\bibnamefont
  {Cramer}}, \bibinfo {author} {\bibfnamefont {C.~M.}\ \bibnamefont {Dawson}},
  \bibinfo {author} {\bibfnamefont {J.}~\bibnamefont {Eisert}},  and\ \bibinfo
  {author} {\bibfnamefont {T.~J.}\ \bibnamefont {Osborne}},\ }\href@noop {}
  {\bibfield  {journal} {\bibinfo  {journal} {Phys. Rev. Lett.}\ }\textbf
  {\bibinfo {volume} {100}},\ \bibinfo {pages} {030602} (\bibinfo {year}
  {2008})}\BibitemShut {NoStop}%
\bibitem{barthel_schollwock_08}%
  \BibitemOpen
  \bibfield  {author} {\bibinfo {author} {\bibfnamefont {T.}~\bibnamefont
  {Barthel}} and\ \bibinfo {author} {\bibfnamefont {U.}~\bibnamefont
  {Schollw\"ock}},\ }\href@noop {} {\bibfield  {journal} {\bibinfo  {journal}
  {Phys. Rev. Lett.}\ }\textbf {\bibinfo {volume} {100}},\ \bibinfo {pages}
  {100601} (\bibinfo {year} {2008})}\BibitemShut {NoStop}%
\bibitem{kollar_eckstein_08}%
  \BibitemOpen
  \bibfield  {author} {\bibinfo {author} {\bibfnamefont {M.}~\bibnamefont
  {Kollar}} and\ \bibinfo {author} {\bibfnamefont {M.}~\bibnamefont
  {Eckstein}},\ }\href@noop {} {\bibfield  {journal} {\bibinfo  {journal}
  {Phys. Rev. A}\ }\textbf {\bibinfo {volume} {78}},\ \bibinfo {pages} {013626}
  (\bibinfo {year} {2008})}\BibitemShut {NoStop}%
\bibitem{cramer_flesch_08b}%
  \BibitemOpen
  \bibfield  {author} {\bibinfo {author} {\bibfnamefont {A.}~\bibnamefont
  {Flesch}}, \bibinfo {author} {\bibfnamefont {M.}~\bibnamefont {Cramer}},
  \bibinfo {author} {\bibfnamefont {I.~P.}\ \bibnamefont {McCulloch}}, \bibinfo
  {author} {\bibfnamefont {U.}~\bibnamefont {Schollw\"ock}},  and\ \bibinfo
  {author} {\bibfnamefont {J.}~\bibnamefont {Eisert}},\ }\href@noop {}
  {\bibfield  {journal} {\bibinfo  {journal} {Phys. Rev. A}\ }\textbf {\bibinfo
  {volume} {78}},\ \bibinfo {pages} {033608} (\bibinfo {year}
  {2008})}\BibitemShut {NoStop}%
\bibitem{rossini_silva_09}%
  \BibitemOpen
  \bibfield  {author} {\bibinfo {author} {\bibfnamefont {D.}~\bibnamefont
  {Rossini}}, \bibinfo {author} {\bibfnamefont {A.}~\bibnamefont {Silva}},
  \bibinfo {author} {\bibfnamefont {G.}~\bibnamefont {Mussardo}},  and\
  \bibinfo {author} {\bibfnamefont {G.~E.}\ \bibnamefont {Santoro}},\
  }\href@noop {} {\bibfield  {journal} {\bibinfo  {journal} {Phys. Rev. Lett.}\
  }\textbf {\bibinfo {volume} {102}},\ \bibinfo {pages} {127204} (\bibinfo
  {year} {2009})}\BibitemShut {NoStop}%
\bibitem{iucci_cazalilla_09}%
  \BibitemOpen
  \bibfield  {author} {\bibinfo {author} {\bibfnamefont {A.}~\bibnamefont
  {Iucci}} and\ \bibinfo {author} {\bibfnamefont {M.~A.}\ \bibnamefont
  {Cazalilla}},\ }\href@noop {} {\bibfield  {journal} {\bibinfo  {journal}
  {Phys. Rev. A}\ }\textbf {\bibinfo {volume} {80}},\ \bibinfo {pages} {063619}
  (\bibinfo {year} {2009})}\BibitemShut {NoStop}%
\bibitem{mossel_caux_10}%
  \BibitemOpen
  \bibfield  {author} {\bibinfo {author} {\bibfnamefont {J.}~\bibnamefont
  {Mossel}} and\ \bibinfo {author} {\bibfnamefont {J.-S.}\ \bibnamefont
  {Caux}},\ }\href@noop {} {\bibfield  {journal} {\bibinfo  {journal} {New J.
  Phys.}\ }\textbf {\bibinfo {volume} {12}},\ \bibinfo {pages} {055028}
  (\bibinfo {year} {2010})}\BibitemShut {NoStop}%
\bibitem{barmettler_punk_10}%
  \BibitemOpen
  \bibfield  {author} {\bibinfo {author} {\bibfnamefont {P.}~\bibnamefont
  {Barmettler}}, \bibinfo {author} {\bibfnamefont {M.}~\bibnamefont {Punk}},
  \bibinfo {author} {\bibfnamefont {V.}~\bibnamefont {Gritsev}}, \bibinfo
  {author} {\bibfnamefont {E.}~\bibnamefont {Demler}},  and\ \bibinfo {author}
  {\bibfnamefont {E.}~\bibnamefont {Altman}},\ }\href@noop {} {\bibfield
  {journal} {\bibinfo  {journal} {New J. Phys.}\ }\textbf {\bibinfo {volume}
  {12}},\ \bibinfo {pages} {055017} (\bibinfo {year} {2010})}\BibitemShut
  {NoStop}%
\bibitem{fioretto_mussardo_10}%
  \BibitemOpen
  \bibfield  {author} {\bibinfo {author} {\bibfnamefont {D.}~\bibnamefont
  {Fioretto}} and\ \bibinfo {author} {\bibfnamefont {G.}~\bibnamefont
  {Mussardo}},\ }\href@noop {} {\bibfield  {journal} {\bibinfo  {journal} {New
  J. Phys.}\ }\textbf {\bibinfo {volume} {12}},\ \bibinfo {pages} {055015}
  (\bibinfo {year} {2010})}\BibitemShut {NoStop}%
\bibitem{cassidy_clark_11}%
  \BibitemOpen
  \bibfield  {author} {\bibinfo {author} {\bibfnamefont {A.~C.}\ \bibnamefont
  {Cassidy}}, \bibinfo {author} {\bibfnamefont {C.~W.}\ \bibnamefont {Clark}},
   and\ \bibinfo {author} {\bibfnamefont {M.}~\bibnamefont {Rigol}},\
  }\href@noop {} {\bibfield  {journal} {\bibinfo  {journal} {Phys. Rev. Lett.}\
  }\textbf {\bibinfo {volume} {106}},\ \bibinfo {pages} {140405} (\bibinfo
  {year} {2011})}\BibitemShut {NoStop}%
\bibitem{kollath_lauchli_07}%
  \BibitemOpen
  \bibfield  {author} {\bibinfo {author} {\bibfnamefont {C.}~\bibnamefont
  {Kollath}}, \bibinfo {author} {\bibfnamefont {A.~M.}\ \bibnamefont
  {L\"auchli}},  and\ \bibinfo {author} {\bibfnamefont {E.}~\bibnamefont
  {Altman}},\ }\href@noop {} {\bibfield  {journal} {\bibinfo  {journal} {Phys.
  Rev. Lett.}\ }\textbf {\bibinfo {volume} {98}},\ \bibinfo {pages} {180601}
  (\bibinfo {year} {2007})}\BibitemShut {NoStop}%
\bibitem{manmana_wessel_07}%
  \BibitemOpen
  \bibfield  {author} {\bibinfo {author} {\bibfnamefont {S.~R.}\ \bibnamefont
  {Manmana}}, \bibinfo {author} {\bibfnamefont {S.}~\bibnamefont {Wessel}},
  \bibinfo {author} {\bibfnamefont {R.~M.}\ \bibnamefont {Noack}},  and\
  \bibinfo {author} {\bibfnamefont {A.}~\bibnamefont {Muramatsu}},\ }\href@noop
  {} {\bibfield  {journal} {\bibinfo  {journal} {Phys. Rev. Lett.}\ }\textbf
  {\bibinfo {volume} {98}},\ \bibinfo {pages} {210405} (\bibinfo {year}
  {2007})}\BibitemShut {NoStop}%
\bibitem{rigol_dunjko_08}%
  \BibitemOpen
  \bibfield  {author} {\bibinfo {author} {\bibfnamefont {M.}~\bibnamefont
  {Rigol}}, \bibinfo {author} {\bibfnamefont {V.}~\bibnamefont {Dunjko}}, 
  and\ \bibinfo {author} {\bibfnamefont {M.}~\bibnamefont {Olshanii}},\
  }\href@noop {} {\bibfield  {journal} {\bibinfo  {journal} {Nature}\ }\textbf
  {\bibinfo {volume} {452}},\ \bibinfo {pages} {854} (\bibinfo {year}
  {2008})}\BibitemShut {NoStop}%
\bibitem{moeckel_kehrein_08}%
  \BibitemOpen
  \bibfield  {author} {\bibinfo {author} {\bibfnamefont {M.}~\bibnamefont
  {Moeckel}} and\ \bibinfo {author} {\bibfnamefont {S.}~\bibnamefont
  {Kehrein}},\ }\href@noop {} {\bibfield  {journal} {\bibinfo  {journal} {Phys.
  Rev. Lett.}\ }\textbf {\bibinfo {volume} {100}},\ \bibinfo {pages} {175702}
  (\bibinfo {year} {2008})}\BibitemShut {NoStop}%
\bibitem{reimann_08}%
  \BibitemOpen
  \bibfield  {author} {\bibinfo {author} {\bibfnamefont {P.}~\bibnamefont
  {Reimann}},\ }\href@noop {} {\bibfield  {journal} {\bibinfo  {journal} {Phys.
  Rev. Lett.}\ }\textbf {\bibinfo {volume} {101}},\ \bibinfo {pages} {190403}
  (\bibinfo {year} {2008})}\BibitemShut {NoStop}%
\bibitem{roux_09}%
  \BibitemOpen
  \bibfield  {author} {\bibinfo {author} {\bibfnamefont {G.}~\bibnamefont
  {Roux}},\ }\href@noop {} {\bibfield  {journal} {\bibinfo  {journal} {Phys.
  Rev. A}\ }\textbf {\bibinfo {volume} {79}},\ \bibinfo {pages} {021608}
  (\bibinfo {year} {2009})}\BibitemShut {NoStop}%
\bibitem{eckstein_kollar_09}%
  \BibitemOpen
  \bibfield  {author} {\bibinfo {author} {\bibfnamefont {M.}~\bibnamefont
  {Eckstein}}, \bibinfo {author} {\bibfnamefont {M.}~\bibnamefont {Kollar}}, 
  and\ \bibinfo {author} {\bibfnamefont {P.}~\bibnamefont {Werner}},\
  }\href@noop {} {\bibfield  {journal} {\bibinfo  {journal} {Phys. Rev. Lett.}\
  }\textbf {\bibinfo {volume} {103}},\ \bibinfo {pages} {056403} (\bibinfo
  {year} {2009})}\BibitemShut {NoStop}%
\bibitem{rigol_09a}%
  \BibitemOpen
  \bibfield  {author} {\bibinfo {author} {\bibfnamefont {M.}~\bibnamefont
  {Rigol}},\ }\href@noop {} {\bibfield  {journal} {\bibinfo  {journal} {Phys.
  Rev. Lett.}\ }\textbf {\bibinfo {volume} {103}},\ \bibinfo {pages} {100403}
  (\bibinfo {year} {2009}{\natexlab{a}})}\BibitemShut {NoStop}%
\bibitem{rigol_09b}%
  \BibitemOpen
  \bibfield  {author} {\bibinfo {author} {\bibfnamefont {M.}~\bibnamefont
  {Rigol}},\ }\href@noop {} {\bibfield  {journal} {\bibinfo  {journal} {Phys.
  Rev. A}\ }\textbf {\bibinfo {volume} {80}},\ \bibinfo {pages} {053607}
  (\bibinfo {year} {2009}{\natexlab{b}})}\BibitemShut {NoStop}%
\bibitem{moeckel_kehrein_09}%
  \BibitemOpen
  \bibfield  {author} {\bibinfo {author} {\bibfnamefont {M.}~\bibnamefont
  {Moeckel}} and\ \bibinfo {author} {\bibfnamefont {S.}~\bibnamefont
  {Kehrein}},\ }\href@noop {} {\bibfield  {journal} {\bibinfo  {journal} {Ann.
  Phys.}\ }\textbf {\bibinfo {volume} {324}},\ \bibinfo {pages} {2146}
  (\bibinfo {year} {2009})}\BibitemShut {NoStop}%
\bibitem{santos_rigol_10STATc}%
  \BibitemOpen
  \bibfield  {author} {\bibinfo {author} {\bibfnamefont {L.~F.}\ \bibnamefont
  {Santos}} and\ \bibinfo {author} {\bibfnamefont {M.}~\bibnamefont {Rigol}},\
  }\href@noop {} {\bibfield  {journal} {\bibinfo  {journal} {Phys. Rev. E}\
  }\textbf {\bibinfo {volume} {81}},\ \bibinfo {pages} {036206} (\bibinfo
  {year} {2010})}\BibitemShut {NoStop}%
\bibitem{roux_10}%
  \BibitemOpen
  \bibfield  {author} {\bibinfo {author} {\bibfnamefont {G.}~\bibnamefont
  {Roux}},\ }\href@noop {} {\bibfield  {journal} {\bibinfo  {journal} {Phys.
  Rev. A}\ }\textbf {\bibinfo {volume} {81}},\ \bibinfo {pages} {053604}
  (\bibinfo {year} {2010})}\BibitemShut {NoStop}%
\bibitem{reimann_10}%
  \BibitemOpen
  \bibfield  {author} {\bibinfo {author} {\bibfnamefont {P.}~\bibnamefont
  {Reimann}},\ }\href@noop {} {\bibfield  {journal} {\bibinfo  {journal} {New
  J. Phys.}\ }\textbf {\bibinfo {volume} {12}},\ \bibinfo {pages} {055027}
  (\bibinfo {year} {2010})}\BibitemShut {NoStop}%
\bibitem{rigol_santos_10}%
  \BibitemOpen
  \bibfield  {author} {\bibinfo {author} {\bibfnamefont {M.}~\bibnamefont
  {Rigol}} and\ \bibinfo {author} {\bibfnamefont {L.~F.}\ \bibnamefont
  {Santos}},\ }\href@noop {} {\bibfield  {journal} {\bibinfo  {journal} {Phys.
  Rev. A}\ }\textbf {\bibinfo {volume} {82}},\ \bibinfo {pages} {011604(R)}
  (\bibinfo {year} {2010})}\BibitemShut {NoStop}%
\bibitem{deutsch_91}%
  \BibitemOpen
  \bibfield  {author} {\bibinfo {author} {\bibfnamefont {J.~M.}\ \bibnamefont
  {Deutsch}},\ }\href@noop {} {\bibfield  {journal} {\bibinfo  {journal} {Phys.
  Rev. A}\ }\textbf {\bibinfo {volume} {43}},\ \bibinfo {pages} {2046}
  (\bibinfo {year} {1991})}\BibitemShut {NoStop}%
\bibitem{srednicki_94}%
  \BibitemOpen
  \bibfield  {author} {\bibinfo {author} {\bibfnamefont {M.}~\bibnamefont
  {Srednicki}},\ }\href@noop {} {\bibfield  {journal} {\bibinfo  {journal}
  {Phys. Rev. E}\ }\textbf {\bibinfo {volume} {50}},\ \bibinfo {pages} {888}
  (\bibinfo {year} {1994})}\BibitemShut {NoStop}%
\bibitem{lieb_shultz_61}%
  \BibitemOpen
  \bibfield  {author} {\bibinfo {author} {\bibfnamefont {E.}~\bibnamefont
  {Lieb}}, \bibinfo {author} {\bibfnamefont {T.}~\bibnamefont {Shultz}},  and\
  \bibinfo {author} {\bibfnamefont {D.}~\bibnamefont {Mattis}},\ }\href@noop {}
  {\bibfield  {journal} {\bibinfo  {journal} {Ann. Phys. (NY)}\ }\textbf
  {\bibinfo {volume} {16}},\ \bibinfo {pages} {406} (\bibinfo {year}
  {1961})}\BibitemShut {NoStop}%
\bibitem{holstein_primakoff_40}%
  \BibitemOpen
  \bibfield  {author} {\bibinfo {author} {\bibfnamefont {T.}~\bibnamefont
  {Holstein}} and\ \bibinfo {author} {\bibfnamefont {H.}~\bibnamefont
  {Primakoff}},\ }\href@noop {} {\bibfield  {journal} {\bibinfo  {journal}
  {Phys. Rev.}\ }\textbf {\bibinfo {volume} {58}},\ \bibinfo {pages} {1098}
  (\bibinfo {year} {1940})}\BibitemShut {NoStop}%
\bibitem{jordan_wigner_28}%
  \BibitemOpen
  \bibfield  {author} {\bibinfo {author} {\bibfnamefont {P.}~\bibnamefont
  {Jordan}} and\ \bibinfo {author} {\bibfnamefont {E.}~\bibnamefont
  {Wigner}},\ }\href@noop {} {\bibfield  {journal} {\bibinfo  {journal} {Z.
  Phys.}\ }\textbf {\bibinfo {volume} {47}},\ \bibinfo {pages} {631} (\bibinfo
  {year} {1928})}\BibitemShut {NoStop}%
\bibitem{rigol_muramatsu_05a}%
  \BibitemOpen
  \bibfield  {author} {\bibinfo {author} {\bibfnamefont {M.}~\bibnamefont
  {Rigol}} and\ \bibinfo {author} {\bibfnamefont {A.}~\bibnamefont
  {Muramatsu}},\ }\href@noop {} {\bibfield  {journal} {\bibinfo  {journal}
  {Phys. Rev. A}\ }\textbf {\bibinfo {volume} {72}},\ \bibinfo {pages} {013604}
  (\bibinfo {year} {2005}{\natexlab{a}})}\BibitemShut {NoStop}%
\bibitem{rigol_muramatsu_05b}%
  \BibitemOpen
  \bibfield  {author} {\bibinfo {author} {\bibfnamefont {M.}~\bibnamefont
  {Rigol}} and\ \bibinfo {author} {\bibfnamefont {A.}~\bibnamefont
  {Muramatsu}},\ }\href@noop {} {\bibfield  {journal} {\bibinfo  {journal}
  {Mod. Phys. Lett.}\ }\textbf {\bibinfo {volume} {19}},\ \bibinfo {pages}
  {861} (\bibinfo {year} {2005}{\natexlab{b}})}\BibitemShut {NoStop}%
\bibitem{rigol_05}%
  \BibitemOpen
  \bibfield  {author} {\bibinfo {author} {\bibfnamefont {M.}~\bibnamefont
  {Rigol}},\ }\href@noop {} {\bibfield  {journal} {\bibinfo  {journal} {Phys.
  Rev. A}\ }\textbf {\bibinfo {volume} {72}},\ \bibinfo {pages} {063607}
  (\bibinfo {year} {2005})}\BibitemShut {NoStop}%
\bibitem{numerical_recipes_07}%
  \BibitemOpen
  \bibfield  {author} {\bibinfo {author} {\bibfnamefont {W.~H.}\ \bibnamefont
  {Press}}, \bibinfo {author} {\bibfnamefont {S.~A.}\ \bibnamefont
  {Teukolsky}}, \bibinfo {author} {\bibfnamefont {W.~T.}\ \bibnamefont
  {Vetterling}},  and\ \bibinfo {author} {\bibfnamefont {B.~P.}\ \bibnamefont
  {Flannery}},\ }\href@noop {} {\emph {\bibinfo {title} {Numerical Recipes: The
  Art of Scientific Computing, Third Edition}}}\ (\bibinfo  {publisher}
  {Cambridge University Press},\ \bibinfo {address} {Cambridge},\ \bibinfo
  {year} {2007})\BibitemShut {NoStop}%
\end{thebibliography}
\end{document}